\documentclass[aps,prl,twocolumn,showpacs,preprintnumbers,amsmath,amssymb]{revtex4}

\usepackage{graphicx}
\usepackage{dcolumn}
\usepackage{bm}

\begin{document}

\preprint{ }

\title{Nonequilibrium electron rings for synchrotron radiation production}

\affiliation{The University of Manchester and Cockcroft Institute, Manchester M13 9PL, United Kingdom}
\affiliation{STFC Daresbury Laboratory and Cockcroft Institute, Daresbury Science and Innovation Campus, Warrington, WA4 4AD, United Kingdom}
\affiliation{The University of Oxford, Oxford, United Kingdom}

\author{Hywel Owen\email{hywel.owen@manchester.ac.uk}} \affiliation{The University of Manchester and Cockcroft Institute, Manchester M13 9PL, United Kingdom}
\author{Peter H. Williams} \affiliation{STFC Daresbury Laboratory and Cockcroft Institute, Daresbury Science and Innovation Campus, Warrington WA4 4AD, United Kingdom}
\author{Scott Stevenson} \affiliation{The University of Oxford, Oxford OX1 2JD, United Kingdom}

\date{\today}

\begin{abstract}
Electron storage rings used for the production of synchrotron radiation (SR) have an output photon brightness that is limited by the equilibrium beam emittance. By using interleaved injection and ejection of bunches from a source with repetition rate greater than 1~kHz, we show that it is practicable to overcome this limit in rings of energy $\sim1$~GeV. Sufficiently short kicker pulse lengths enable effective currents of many milliamperes, which can deliver a significant flux of diffraction-limited soft X-ray photons. Thus, either existing SR facilities may be adapted for non-equilibrium operation, or the technique applied to construct SR rings smaller than their storage ring equivalent.
\end{abstract}

\maketitle

The storage ring - developed from the synchrotron \cite{McMillan:1945p6736,Goward:1946p6746} - overcomes the relativistic limitation of fixed-target particle physics experiments by colliding counter-rotating beams of particles \cite{ONeill:1956p6660,ONeill:1963p6661}. The emitted synchrotron radiation (SR) \cite{Schwinger:1949p5463} from cycling electron synchrotrons was already used parasitically from the 1970s for techniques such as X-ray crystallography \cite{Rowe:1973p6818,Worgan:1982p6803,Suller:1978p6804}. Second- and third-generation facilities \cite{Thompson:1980p6811} use storage rings; third-generation rings provide magnet-free straight sections with zero (or small) dispersion \cite{Thompson:1982p6815,Thompson:1983p6810}, and may accommodate periodic magnetic insertion devices (IDs) that induce strong (and sometimes coherent) photon emission at selected wavelengths. The use of electron storage rings for SR blossomed in the 1980s and is now a mature field with around sixty facilities conducting a myriad of experiments \cite{Laclare:2001p6797,Bilderback:2005p6802}. Electron beam quality is characterised by the emittances $\epsilon_x$ and $\epsilon_y$, where, for a single dipole radius $\rho$, $\epsilon_x=55 \hbar \gamma^2 \langle H_x\rangle_s /32 \sqrt{3} m_e c J_x \rho$ is determined by the energy (via $\gamma$) and by the magnetic lattice via $\left<H_x\right>_s=\left<\beta_x \eta_x'^2+2\alpha_x\eta_x\eta_x'+\gamma_x\eta_x^2\right>_s$ ($J_x$  is the horizontal damping partition number). Demand for better resolution pushes ring design to ever-smaller $\epsilon_x$ using lattices that minimise $\left<H_x\right>_s$ (and thereby the equilibrium between radiation damping and quantum excitation), for example the `Theoretical Minimum Emittance' (TME) lattice \cite{Lee:2010p6491,Lee:2003p6748,Lee:2004p6749,Wang:2009p4572} in which only a small fraction of the circumference $C$ may be used for IDs. $\epsilon_y$ is coupled to $\epsilon_x$ mainly by ring magnet roll misalignments \cite{Loulergue:1998p797,ref30}: originally 10\% for the European Synchrotron Radiation Facility (ESRF) \cite{Witte:1990p6788,Anonymous:1987p6787}, coupling values better than 0.1\% are achieved today \cite{Bartolini:2008p6789}, smaller than the diffraction limit for many IDs. One may also increase $C$ to incorporate more dipoles to minimise $<\eta_x>$ in them \cite{Ropert:2000p6801,Elleaume:2003p6798,Tsumaki:2006p6796}, conceptually simple but expensive and yielding only modest gains: it is therefore unlikely that rings much larger than the $\sim 1$ km APS, ESRF and SPRING-8 will be built (PETRA-III is a special case of a large pre-existing ring for particle physics \cite{Balewski:2011p6800}).

Fourth-generation SR facilities overcome some limitations of storage rings, particularly the linac-driven free-electron laser (FEL) \cite{Fuoss:1988p6813,Winick:1994p6819,Arthur:1995p6820} first demonstrated at X-ray wavelengths at the LCLS \cite{Emma:2009p6826,Ratner:2009p6825}. Whilst FELs may provide enormous peak brightness, their average flux is rather modest in comparison with storage rings due to the limitations in either gun or linac macropulse repetition rate. Average flux may be increased by using superconducting cavities and high-current electron injectors: the energy-recovery linac (ERL) uses both to provide quasi-continuous bunch trains whilst alleviating power requirements in the cavities: each electron's energy is recycled as the returned bunches are decelerated, and passed to new electrons \cite{Neil:2002ci,Quinn:2005p4152}. JLab has demonstrated 10~kW of FEL power at 100~MeV electron energy in a 10~mA ERL \cite{Neil:2000dt,Neil:2006cn}.

\begin{figure}
    \includegraphics[width=80mm]{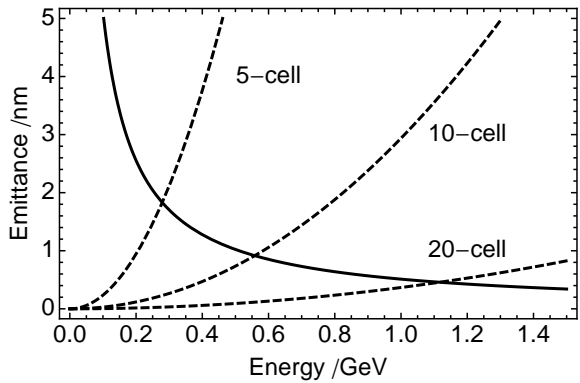}
    \caption{\label{tmeemitvariation} Variation of circulating beam emittance in a storage ring versus energy, comparing an injector (with normalised emittance $\epsilon_n=10^{-6}$~m) with equilibrium emittances for storage rings composed from TME double-bend achromats of 5, 10 and 20 cells (dashed lines).}
\end{figure}

Low-energy ERLs may circulate large currents without the strong intrabeam scattering (IBS) that would limit the emittance of a storage ring at the same energy \cite{Hoffstaetter:2006p4153,Smith:2006p4151}. At sufficiently large electron energies ERLs gain an emittance advantage over that possible from a storage ring. We recall that in the absence of dilution the emittance $\epsilon = \epsilon_n/\beta \gamma$ reduces adiabatically under acceleration. In Figure~\ref{tmeemitvariation} we compare the achievable emittances from a typical modern electron injector to TME lattices of differing size. This emittance advantage has led to several proposed facilities (e.g. \cite{4GLSCDR:2006}), including the conversion of existing storage rings \cite{cornellerl,apserl}. However, since each electron only contributes once to the effective current the charge passing through the accelerator is very high and beam loss control will be difficult. Several proposals assume tens of MWs of circulating beam power and losses kept to a few W/m: for example, a 6~GeV ERL circulating 100~mA through 1~km deposits 600~kW continuously if losses are kept to typical values $\sim$0.1\%, whereas the equivalent storage ring would deposit only 2000~J if a fill were completely lost. An ERL cavity injection scheme may be used to circulate each injected bunch four times instead of just once \cite{Nakamura:2008p6786}, but extending that concept is very challenging.

All present-day electron storage rings circulate bunches where classical SR damping and quantum excitation give an equilibrium emittance $\epsilon_{\mathrm{eq} (x,y)}$, the stored charge either accumulated over repeated bunch injection or transferred in one train from another storage ring. The injected emittances $\epsilon_{\mathrm{inj} (x,y)}$ are usually larger than $\epsilon_{\mathrm{eq} (x,y)}$, particularly in the vertical plane, although this need not be the case \cite{slsinjector}. At energies above $\sim 1$~GeV the damping time $\tau_{x,y} = 3m_e^3c^5 C\rho/(2\pi r_e J_{x,y}E^3)$ falls to a few milliseconds: linear collider damping rings use this to take a (relatively) large injector emittance and reduce it; bunches are left in just long enough to sufficiently damp before they are passed to a downstream linac.

We propose the opposite process, in which low-emittance bunches are repeatedly injected into a ring and grow larger: they are ejected and replaced with new ones well before equilibrium is obtained, so that the effective emittance is essentially $\epsilon_{\mathrm{inj}}$. This process may be carried out in any of the radio-frequency (RF) buckets, limited by the injection rate. This concept has been proposed to improve Compton photon production \cite{Yu:2009p6791} and for electron cooling \cite{ecool}, but not yet proposed for the production of SR. At large enough energies $\tau_{x,y}$ will be too short to allow repeated injection at a feasible rate. However, we here show an energy range exists between a lower-energy emittance crossover point and the damping rate limit where there is an emittance advantage with reasonable average output flux.

\begin{table}
\caption{\label{tab:max23properties}Principal properties of the MAX-II, MAX-III and Super-ACO storage rings when operated in equilibrium.}
\begin{ruledtabular}
\begin{tabular}{llll}
&  MAX-II & MAX-III & Super-ACO \\
\hline
Max. Energy (GeV) & 1.5 & 0.7 & 0.8\\
Design current (mA) & 200 & 250 & 400 \\
C (m) & 90 & 36 & 72\\
$\rho$ (m) & 3.33 & 3.036 & 1.7\\
$T_{\mathrm{rev}}$ (ns) & 300 & 120 & 240\\
$\epsilon_{\mathrm{eq},(x)}$ (nm) & 8.9 & 12.8 & 38.0 \\
$\tau_{x,y}$ (ms) & 6.7 & 24 & 18 \\
$\sigma_{t}$ ($1\sigma$, ps) & 53 & 89 & 90 \\
$\sigma_{E}$ ($1\sigma$) $(/10^{-4})$ & 7.1 & 8.6 & 5.3 \\
$J_x$ & 1.0 & 2.4 & 1.0 \\
\end{tabular}
\end{ruledtabular}
\end{table}

We illustrate this non-equilibrium (NEQ) approach using the designs of the existing Swedish MAX-II \cite{Sjostrom:2007p6753,leblanc:1996} and MAX-III storage rings \cite{Sjostrom:2009p6760} and the French Super-ACO ring \cite{superacoring,superacoring2}, which in normal operation have the properties given in Table \ref{tab:max23properties}; the energy of each ring may be readily lowered to reduce $\epsilon_{\mathrm{eq} (x)}$ and increase $\tau_{x,y}$. We assume full-energy injection of bunches with charge $q=1$~nC is possible at a frequency $f_l$ up to 10~kHz, from a suitable injector with $\epsilon_n=10^{-6}$~m: several FEL proposals indicate these parameters are accessible \cite{apexgun,nlscdr}. We consider the use of fast bunch-by-bunch injection and extraction as proposed for collider damping rings, for example the 0.73~m DAF$\Phi$NE and 1.4~m KEK kicker designs where short enough rise/fall times (12.4~ns and 6~ns) and peak-to-peak (p-p) stability of 0.07\% are possible with deflection angles at 1~GeV of 2.6~mrad and 3.0~mrad respectively \cite{Alesini:2006p6755,Naito:2007p6756}; continuous repetition rates over 1~kHz have been proposed for other fast kickers \cite{nlscdr,PlacidiPappas}. Over straight lengths of a few metres the total kick angle from several kickers ($> 5$~mrad for $E<1$~GeV) is sufficient to translate the beam at least 10~mm from an injection septum, thereby providing ample Touschek and quantum lifetime for the circulating beam. 

There is no significant radiation damping, since by definition electrons are resident for a time smaller than $\tau_{x,y}$. Emittance dilution due to injection steering errors will persist, and may be estimated as $\Delta\epsilon / \epsilon = (\Delta x^2 + (\beta_0 \Delta x'+\alpha_0 \Delta x)^2)/2\beta_0\epsilon$. Dilution is small at energies below 1~GeV for achievable p-p stabilities and may be suppressed by reducing $\beta_0$ at the injection point; other dilution effects such as optical mismatch and steering fluctuations are expected to be small \cite{Fischer:1997p6757}.

The low energy and large bunch charges mean that IBS is important \cite{Kubo:2005p6794}. Growth rate estimates using \textsc{elegant} \cite{Borland2000} indicate initial characteristic times $\tau_{\mathrm{IBS}} \lesssim 1$~ms for injected bunch lengths $\sigma_t\sim1$~ps; $\tau_{\mathrm{IBS}}$ varies linearly with $\sigma_t$, and so may be made greater than $\tau_{x,y}$ by increasing $\sigma_t$ from 1~ps to the equilibrium stored lengths of either 50 or 90~ps. Large injected $\sigma_t$ may be achieved either with linac RF phasing or with bunch shear from a magnetic chicane. For large $\sigma_t$ the variation in deflection angle over the kicker pulse is still small, and even at $\sigma_t\sim$~90~ps the dilution effect on the circulating emittance is less than 1\%. Interleaved injection/extraction is therefore feasible, but the alternative scheme of repeatedly injecting bunch trains is not, since the much longer required kicker pulse will vary significantly in amplitude over the train. We note that other collective effects such as resistive wall instabilities, microbunching and so on will either be small or controllable with feedback \cite{Hock:2009p5692}.

\begin{figure}
    \includegraphics[width=80mm]{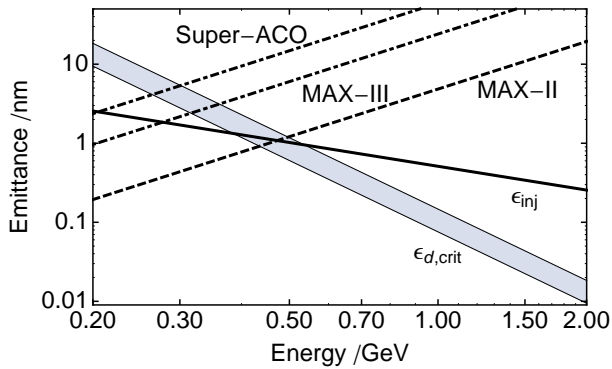}
    \caption{\label{maxemitvariation} Variation of emittance with energy, comparing an injector (with normalised emittance $\epsilon_n=10^{-6}$ m) with the equilibrium emittance of the MAX-II, MAX-III and Super-ACO storage rings. Also shown is the range of the diffraction-limited emittance $\epsilon_{d,\mathrm{crit}}$ for bending radii $\rho$ from 1.7~m to 3.33~m.}
\end{figure}

The number of bunches $n_b$ that may be circulated in NEQ mode is limited by the kicker rise/fall time rather than by $f_l$, as it is in damping rings. For $q=1$~nC the total current that may be circulated is simply $I=q/\tau_b$ where $\tau_b$ is the possible bunch spacing, i.e. $I=8.3$~mA for 12~ns spacing (6~ns kicker rise and fall time), with $n_b=T_{\mathrm{rev}}/\tau_b$; there is an insignificant reduction in effective $I$ due to the time taken to replace each bunch (which is smaller than $T_{\mathrm{rev}}$). The residence time of each bunch in the ring is determined by the repetition rate $f_l$ of the injector as $\tau_r=n_b/f_l$. With a 1~kHz injector rate the residence times are long enough that IBS would determine the emittance, but at $f_l=10$~kHz $\tau_r$ is small enough that no significant emittance growth occurs. Table \ref{tab:basicperformance} gives example operating energies for each of the considered rings, and shows that a significant emittance advantage may be obtained using NEQ operation. 

\begin{table*}
\caption{\label{tab:basicperformance}Non-equilibrium operation of MAX-II, MAX-III and Super-ACO, assuming an injected bunch length equal to the natural bunch length at full energy. The beam current in all cases is 8.3~mA, limited by the 12~ns bunch spacing.}
\begin{ruledtabular}
\begin{tabular}{lllllll}
 &  MAX-II & & MAX-III & & Super-ACO &\\
\hline
Energy (GeV) & 0.7 & 1.0 & 0.5 & 0.7 & 0.5 & 0.8\\
\hline
$n_b$ & 25 & 25 & 10 & 10 & 20 & 20 \\
$\tau_r$ (ms) for $f_l=10$~kHz & 2.5 & 2.5 & 1.0 & 1.0 & 2.0 & 2.0 \\
$\tau_{x,y}$ (ms) & 66 & 23 & 13,32 & 4.8,11.5 & 74 & 18 \\
$\tau_{\mathrm{IBS}x,init}$ (ms) & 13 & 22 & 2.0 & 3.8 & 4.8 & 11.7 \\
$\epsilon_{\mathrm{eq},x}$  (nm) & 2.73 & 3.95 & 6.5 & 12.8 & 15.0 & 38.0 \\
$\epsilon_{\mathrm{inj},x}$, $\epsilon_{\mathrm{inj},y}$ (nm) & 0.730 & 0.511 & 1.28 & 1.02 & 1.02 & 0.639 \\
$\epsilon_{d,\mathrm{crit}}$ (nm) & 0.432 & 0.148 & 2.11 & 1.08 & 0.604 & 0.138 \\
$\Delta\epsilon /\epsilon$ (x,y) from p-p stability (\%) & 16,2.7 & 22,3.8 & 1.0,0.7 & 1.4,1.0 & 5.0,5.8 & 8.0,9.4 \\
Dump power (kW) for $f_l=10$~kHz & 7 & 10 & 5 & 7 & 5 & 8 \\
\end{tabular}
\end{ruledtabular}
\end{table*}

IDs will deliver photons comparable to the critical wavelength $\lambda_c=4\pi\rho/3\gamma^3$ of the main ring dipoles: we may therefore use the diffraction-limited emittance at $\lambda_c$, given as $\epsilon_{d,\mathrm{crit}}=\rho/3\gamma^3$, as an estimate of the useful emittance from NEQ operation; this is compared to the equilibrium and NEQ emittances in Figure~\ref{maxemitvariation}. Equating $\epsilon_{d,\mathrm{crit}}$ with $\epsilon_{\mathrm{inj}}$ we obtain the electron energy at which NEQ operation gives an advantage, $E\simeq m_e c^2 \sqrt{\rho/3 \epsilon_n}$, around 0.4~GeV to 0.7~GeV for typical dipole radii. The upper limit on $E$ is set by ensuring $\tau_r \ll \tau_x$, in other words that  $f_l \gg 2\pi r_e JE^3/3m_e^3 c^6 \rho \tau_b$; this limit is about 1.5~GeV to 2.5~GeV depending on $\rho$, and is similar to the energy limit from kicker emittance dilution. The beam power deposited at the dump is simply that from the injector, and is readily manageable even at $f_l=10$~kHz \cite{4GLSCDR:2006,nlscdr,cebafdumps,xfeldump,nlsdumps}: the beam power is therefore limited by the kicker rise/fall time. A significant beam power saving is obtained over an equivalent ERL, with no emittance penalty.  Compared to an equilibrium storage ring, the horizontal emittance is reduced significantly with NEQ operation and may be brought close to $\epsilon_{d,\mathrm{crit}}$; the equilibrium vertical emittance is typically smaller than $\epsilon_{d,\mathrm{crit}}$, so the increase from NEQ operation incurs no penalty for most IDs. The range of output photon energies over which NEQ operation gives an emittance advantage is from about $\sim 0.1$~keV to $\sim 2$~keV. We suggest that NEQ rings of simple optical design and small $\rho$ may be an inexpensive way to add spontaneous photon beamlines to a linac that drives a soft X-ray or higher energy FEL. A small circumference lattice can give NEQ emittances similar to much larger rings operated in equilibrium, whilst still delivering simultaneous photons with significant flux to numerous IDs and their associated experiments.

This work was supported in part by the Science and Technology Facilities Council. We would like to thank Jim Clarke and James Jones of STFC Daresbury Laboratory, and Sara Thorin of MAX-Lab, for technical information and useful discussions.

\bibliographystyle{apsrevhacked}

\end{document}